\documentclass[twocolumn]{aastex6}
\usepackage{appendix}
\usepackage{color}
\usepackage{amsmath,amssymb,bm, graphicx,xspace,multirow}
\usepackage{epsfig}
\usepackage{lineno}

\def\aap{Astronomy and Astrophysics}

\def\apjl{ApJL}

\def\apjs{ApJS}

\begin{document}

\title{ PeV emission of the Crab Nebula:  constraints on the proton content in pulsar wind and implications}

\author{Ruo-Yu Liu\altaffilmark{1,2}, Xiang-Yu Wang \altaffilmark{1,2}}

\altaffiltext{1}{School of Astronomy and Space Science, Nanjing University, Nanjing 210023, China; E-mail: \textcolor{blue}{ryliu@nju.edu.cn; xywang@nju.edu.cn}}
\altaffiltext{2}{Key laboratory of Modern Astronomy and Astrophysics (Nanjing University), Ministry of Education, Nanjing 210023, China}

\begin{abstract}
Recently, two photons with energy of  about 1\,PeV  have been detected by LHAASO from the Crab nebula, opening an ultra-high energy window for studying the pulsar wind nebulae (PWNe). Remarkably, the LHAASO spectrum at the highest-energy end shows
a possible hardening, which could indicate the presence of a new component. A two-component scenario with a main electron component and a secondary proton component  has been proposed to explain the whole spectrum of the Crab Nebula, requiring a proton energy of $10^{46}-10^{47}{\rm ergs}$ remaining in the present Crab Nebula.  In this paper, we study the energy content of relativistic protons in pulsar winds  with the LHAASO data of the Crab Nebula, considering the effect of diffusive escape of relativistic protons.  Depending on the extent of the escape of relativistic protons, the total energy of protons lost in the pulsar wind could be 10-100 times larger than that remaining in the nebula presently.  We find that the current LHAASO data allows up to $(10-50)\%$ of the spindown energy of pulsars being converted into relativistic protons. The escaping protons from PWNe  could make a considerable contribution to the cosmic-ray  flux of 10-100 PeV. We also discuss the leptonic scenario for the possible spectral hardening at PeV energies.

\end{abstract}


\section{Introduction}
A rotation-powered pulsar converts most of its rotational energy losses into a highly relativistic magnetized wind \citep[e.g.][]{GJ69,KC84a}. The collision of the pulsar wind with the ambient supernova (SN) ejecta and/or interstellar material (ISM) results in a termination shock  and creates a pulsar wind nebula (PWN), a
region of up to tens of pc, filled with relativistic electrons and positrons \citep{Rees74, Reynolds84}. It is widely believed that lower energy radiation in the PWNe is produced by the synchrotron radiation of leptons in the magnetic
field and the higher energy
part of the spectrum is produced by leptons in the inverse Compton scattering of the ambient photons \citep[e.g.][]{KC84b, deJager92, Atoyan96}

The Crab nebula is a unique representative of PWNe,  characterized by a very broad spectral energy distribution (SED) that  spans over 21 decades, from MHz radio wavelengths to ultra high-energy (UHE, $E>100\,$TeV)
gamma-rays. It is powered by the most energetic pulsar PSR~J0534+2200 (or the Crab pulsar) found in our Galaxy with a current spindown luminosity of $L_s=4.5\times10^{38}\rm erg~s^{-1}$. The pulsar has a characteristic age of $\tau_c=1260$\,years while the record in Chinese chronicles shows its true age to be 967\,years \citep{Lundmark1921}. In the Crab nebula, several photon fields serve as targets for the inverse Compton (IC) radiation of electrons \citep{Atoyan96}. Three
dominant IC components are contributed by the far-infrared
(FIR), cosmic microwave background (CMB) radiation , and synchrotron photons. Up-scattering of
synchrotron photons through the synchrotron-self-Compton
(SSC) channel provides the major contribution at TeV energies. The synchrotron target is however characterized by
a relatively high photon energy, therefore in the UHE band
the SSC process is significantly suppressed because of the
Klein-Nishina (KN) effect and the IC scattering process of CMB photons is the dominant component.

The remarkable discovery of PeV photons from Crab Nebula recently by the Large High Altitude Air Shower Observatory (LHAASO) is the first step towards the opening of the PeV window in the cosmic electromagnetic spectrum \citep{LHAASO21_sci,LHAASO21_nat}.
Although the current data of of Crab nebula measured by LHAASO
is consistent with a single log-parabola type spectrum with a photon index close to $-3.7$ at PeV energy, the highest-energy spectrum shows a possible hardening, implying the existence of a second spectral component of either leptonic or hadronic origin as discussed in \citet{LHAASO21_sci}. A two-component scenario with a main electron component and a secondary proton component  has been proposed to explain the whole spectrum of the Crab Nebula\citep{LHAASO21_sci}.
In fact, it has been suggested that a fraction of the spin-down power
of pulsars can be converted into a wind of nuclei \citep{Hoshino92, Arons94, Gallant94}. Nuclei can be  accelerated in pulsar magnetospheres as discussed by \citet{ChengKS86, Bednarek97}. These nuclei may suffer partial photo-disintegration in the non-thermal radiation fields of the pulsar's outer magnetosphere.  The products (protons and neutrons) of the photo-disintegration and surviving heavier nuclei are injected into the PWN, and then interact with the ambient matter.  Thus, $\gamma$-rays are produced via inelastic collisions, and this hadronic origin of gamma-rays has been suggested as an alternative or
an additional component to the IC gamma-rays in the literature \citep[see e.g.][]{ChengKS90, Atoyan96, Aharonian98, Bednarek97, Horns06, YangX09,ZhangL09, LiH10}.

Recently, \citet{ZhangX20} obtained an upper limit of 0.5\% on the energy fraction of relativistic protons contained in the Crab Nebula by modelling the broadband spectrum including the data of the Tibet AS$\gamma$ experiment \citep{ASgamma19}. It may be worth noting that, without considering the escape of protons from the nebula, the constraint does not reflect the true fraction of the energy channelled into relativistic protons from the pulsar wind. \citet{LHAASO21_sci} found that to account for PeV emission via the hadronic process, it needs a  power of $10^{36}\rm erg~s^{-1}$ for 10\,PeV protons  and an order of
magnitude more for a broad $E^{-2}$ spectrum assuming the most effective confinement for protons. This power is about 0.2-2\% of the pulsar's current spindown power.

In this paper, we will study the proton content in pulsar wind using the LHAASO data of the Crab Nebula and considering the diffusive escape effect. The rest of the paper is organized as follows. We study the constraint on the proton content in the Crab Nebula  in Section 2. In Section 3, we will discuss the implications of the result and give a summary in Section 4.


\section{The effect of diffusive escape on the hadronic emission}
The possible hardening of the Crab Nebula spectrum at PeV energies could, in principle, be either leptonic or hadronic. In this section, we will focus on the hadronic scenario, while a brief discussion on the leptonic scenario is presented in the discussion section.

In the hadronic scenario, the PeV photons from the Crab nebula arise from  $pp$ collisions with the matter in the nebula. The energy of protons that produce $E_\gamma=1\,$PeV is $E_p=10E_\gamma=10\,$PeV.  The
collision loss time for relativistic protons interacting with the matter in the nebula is
\begin{equation}
t_{\rm pp}=\frac{1}{\xi_{\rm pp} n \sigma_{\rm pp} c}=10^7{\rm yr} \left(\frac{n}{\rm 10 cm^{-3}}\right)^{-1} \left(\frac{\xi_{\rm pp}}{0.2}\right)^{-1},
\end{equation}
where $n$ is the average gas density in the nebula, $\sigma_{\rm pp}$ is the cross section for proton-proton interaction and $\xi_{\rm pp}$ is the inelasticity of pion production. The mass in the Crab
filaments is estimated to be $7.2\pm0.5 M_\odot$ \citep{Owen15}, so the gas density is $10 \,{\rm cm^{-3}}$ for a volume of $30 \,{\rm pc^3}$.
The luminosity of the PeV gamma-ray emission in the energy range of 0.5-1.1 PeV is about $L_\gamma\simeq 5\times10^{31}{\rm erg s^{-1}}$ \citep{LHAASO21_sci}. Then, the  energy of protons producing these PeV gamma-rays is about
\begin{equation}
W_{p}\simeq L_\gamma t_{pp}=1.5\times10^{46}{\rm erg} \left(\frac{n}{\rm 10 cm^{-3}}\right)^{-1} .
\end{equation}
Assuming a $E^{-2}$ spectrum for protons, the total energy in relativistic protons will be larger by a factor of several. This energy is only a very small fraction of  the total spindown energy of the Crab pulsar, which is about $10^{49}{\rm erg}$.
Note, however, that $W_p$ is the energy of protons that remained in the Crab nebula at present. The real energy of protons that were injected into the nebula could be much larger, since  protons/nuclei injected into the nebula at earlier time could have escaped out of the nebula.

The modelling of PWNe as the result of the magnetohydrodynamic (MHD) downstream flow from a shocked, relativistic pulsar wind has been successful in reproducing many features of the nebulae observed close to the central pulsars \citep{KC84a}. In the model, the particles are advected with a toroidal magnetic field. Cross field scattering of particles is expected to be small \citep[e.g.][]{deJager09}, so that diffusion of particles can be neglected. However, although the core region has a clear toroidal magnetic structure, the outer nebula has a complex structure that includes a radial component to the magnetic field.
In the Crab, the prominent toroidal structure observed at X-ray and optical wavelengths extends only to 40'' from the pulsar \citep{Weisskopf00}, while the nebular radius is 200'' (corresponding to a physical size of approximately $r_{\rm crab}=2\,$pc at a nominal distance of 2\,kpc). In addition, the radial variation of spectral index due to synchrotron losses is smoother than expected in the MHD flow model \citep{Tang12}. Thus, diffusion should be the dominant process for particle transport in the outer nebula. Indeed, \citet{Tang12} found that
the diffusion models can reproduce the basic data on the nebular size and the spectral index variation of some PWNe including the Crab nebula.

\begin{figure}[t]
\centering
\includegraphics[width=1\columnwidth]{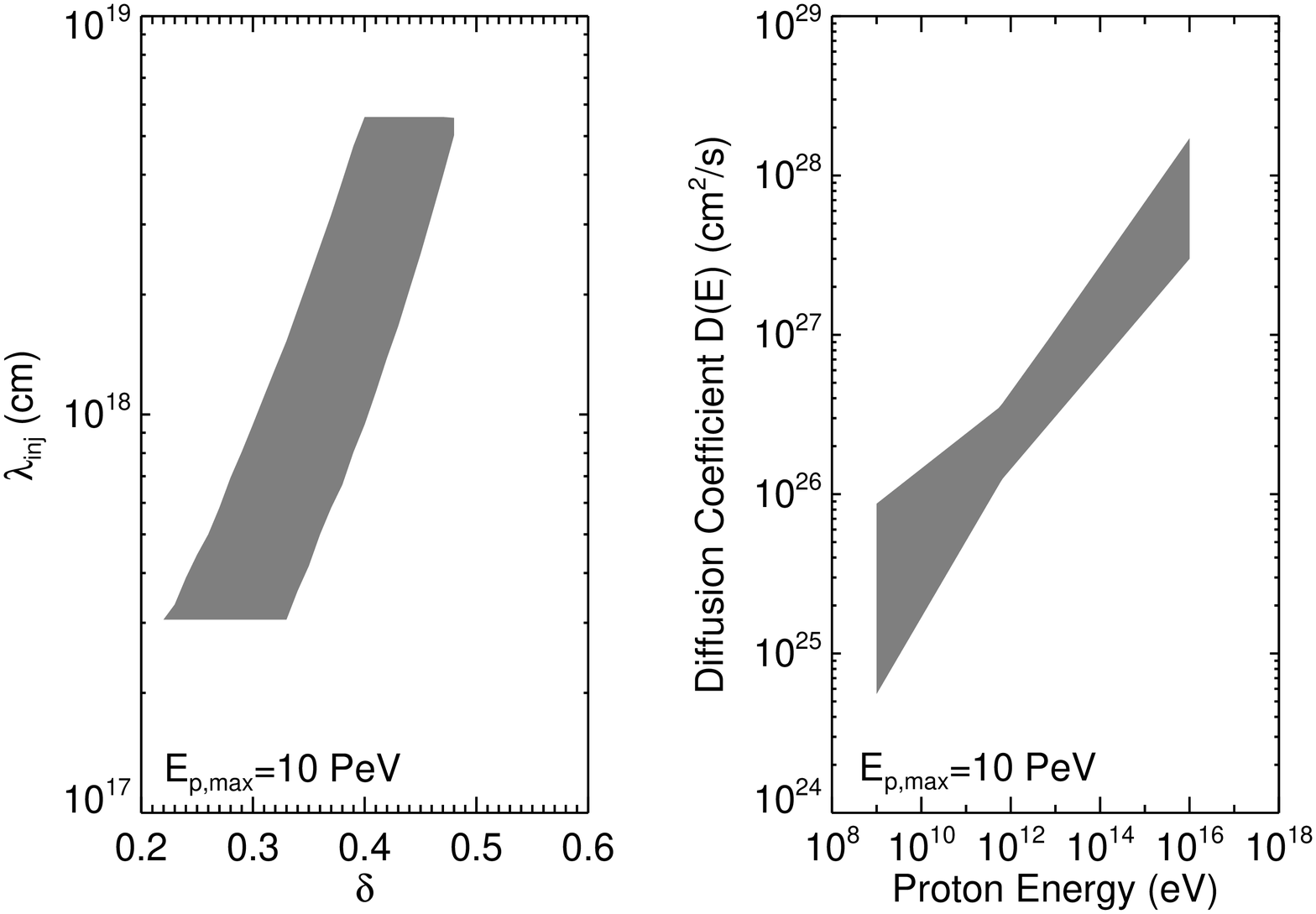}
\includegraphics[width=1\columnwidth]{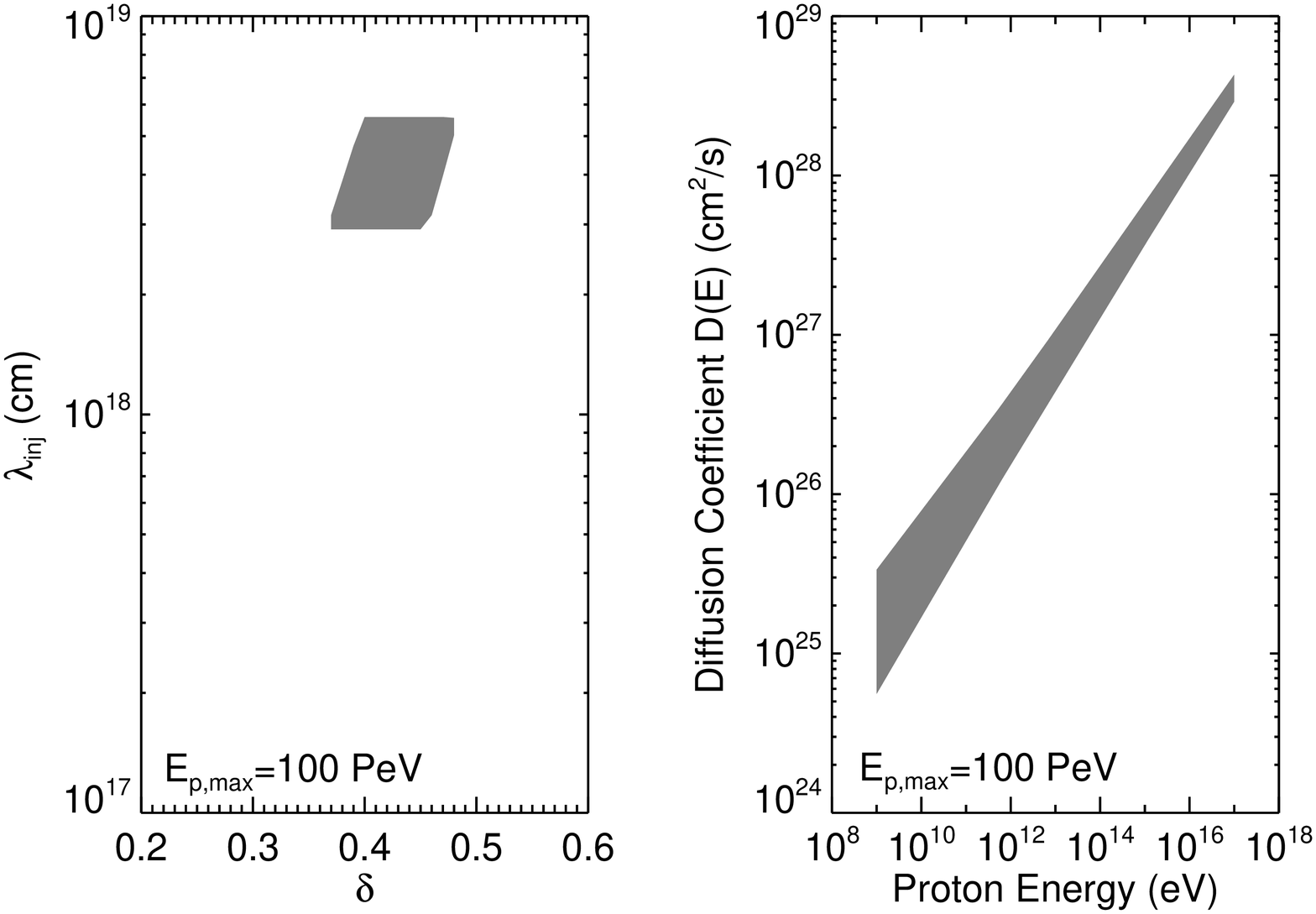}
\caption{{\bf Left:} Allowable parameter space for the index of the diffusion coefficient ($\delta$) and the turbulence injection scale ($\lambda_{\rm inj}$). {\bf Right}: Corresponding diffusion coefficient as a function of particle energy. The upper panels are for $E_{p,\rm max}=10^{16}\,$eV while the lower panels are for $E_{p,\rm max}=10^{17}\,$eV.}
\label{fig:diff}
\end{figure}

The diffusive escape time from the Crab Nebula
with an effective radius $R$ is given by $t_{\rm diff}=R^2/[4D(E)]$ assuming isotropic diffusion, where $D(E)$ is the diffusion coefficient. \citet{Tang12} obtained a diffusion coefficient of $D_0=2.4\times 10^{26}{\rm cm^2 s^{-1}}$ at the energy of $E=0.6 {\rm TeV}$. The diffusion coefficient at 10\,PeV is, however, not known. The particle gyroradius for 10\,PeV protons is $r_g=E/eB=3\times10^{17}{\rm cm} \left(E/10 {\rm PeV}\right)\left(B/100\mu {\rm G}\right)^{-1}$, which is larger than the mean free path ($\simeq 3D_0/c$) of particles at low energies, invalidating the constant diffusion coefficient assumption. We therefore employ a general form for the diffusion coefficient, $D(E)\propto \left(E/0.6{\rm TeV}\right)^{\delta}$. The parameter $\delta$ depends on the power spectrum of the turbulence (i.e., $W(k)\propto k^{\delta-2}$ with $k$ being the wave number) and is usually considered to be $0-1$ in literature, with $\delta=0$ representing the hard-sphere approximation while $\delta=1$ representing the Bohm-type approximation for the turbulence. The CR diffusion is due to the scattering by the plasma wave with wavelength ($\lambda \sim 1/k$) comparable to the gyroradius of the CR. The diffusion coefficient can be given by $D(E)=(c/3)(r_g/kW(k))$ where $kW(k)$ representing the ratio of turbulence energy density at scale $r_g\simeq 1/k$ to the total magnetic energy density. Assuming a strong turbulence with $kW(k)\sim 1$ at the turbulence injection scale $\lambda_{\rm inj}$, we can rewrite the diffusion coefficient to be $D(E)=(c/3)\lambda_{\rm inj}\left(E/\lambda_{\rm inj}eB\right)^\delta$. Note that the injection scale of the turbulence $\lambda_{\rm inj}$ cannot exceed the size of the nebula $r_{\rm crab}=2\,$pc while it should not be smaller than the gyroradius of the maximum proton energy considered here. Such a condition can put a constraint on the value of $\delta$, if we further require the employed diffusion coefficient to be consistent with the inferred one by \citet{Tang12} to certain extent, i.e., $D_0/2<D(0.6\rm TeV)<3D_0/2$ given a 50\% error range. The allowable parameter space for $\lambda_{\rm inj}$ and $\delta$ is shown in Fig.~\ref{fig:diff} for $E_{p,\rm max}=10\,$PeV and 100\,PeV respectively. Also shown are the corresponding diffusion coefficients. The magnetic field is taken to be $B=112\,\mu$G which is the best-fit parameter found by \citet{LHAASO21_sci}.   Note that the mean free path of particle, $\sim 3D/c$, is $\lesssim 1\,$pc,  which is smaller than $r_{\rm crab}$ even for 100\,PeV in the considered scenario here, and hence a diffusion description for the particle transport in the nebula is roughly valid up to 100\,PeV. A more strict treatment should include the ballistic propagation of the highest-energy particle in the core region of the nebula (see e.g. \citealt{Aloisio09, Prosekin15}), which may change the expected gamma-ray flux around 10\,PeV.

For an instant injection of particle, the spatial distribution of particle density at an epoch $t$ after the injection is proportional to $(D(E)t)^{-3/2}\exp\left(-r^2/4D(E)t\right)$ assuming spherically symmetric diffusion \citep{Atoyan95}. This means that the particle distribution approximately keeps constant at $r<r_{\rm diff}\equiv 2\left[D(E)t\right]^{1/2}$ whereas declines quickly at $r>r_{\rm diff}$. Therefore, the fraction of particles remaining inside the nebula approximately reads $f_p(E_p,t)={\rm min}[1,(r_{\rm crab}/r_{\rm diff})^3]$. We assume that the CR injection luminosity is a fraction of $\eta_p$ of the spindown luminosity of the pulsar which evolves with time as $(1+t/\tau_0)^{-\sigma}$ \citep{Pacini73}, where $\tau_0\simeq 680\,$yr is the initial spindown timescale and $\sigma=2.33$ given Crab pulsar's braking index being 2.5 \citep{Lyne88}. The cumulative CR protons that are still confined inside the nebula can be given by
\begin{equation}\label{eq:npspec}
 \frac{dN_p}{dE_p}=\int_0^{\tau_{\rm crab}} \frac{Q_0(E_p)f_p(E_p,t)}{\left[1+(\tau_{\rm crab}-t)/\tau_0\right]^{\sigma}}dt
 \end{equation}
with $\tau_{\rm crab}=967\,$yr is the age of Crab pulsar at today. $Q_0(E_p)=N_0E_p^{-\alpha_p}\exp(-E_p/E_{p,\rm max})$ is the proton injection spectrum where $N_0$ is the normalization factor. The value of $N_0$ can be found by $\int E_pQ_0(E_p)dE_p=\eta_pL_{s,0}$ with $L_{s,0}=L_s(1+\tau_{\rm crab}/\tau_0)^{\sigma}$ being the initial spindown luminosity of the pulsar.

We can find a critical energy $E_c$ for protons by $\tau_{\rm crab}=r_{\rm crab}^2/4D(E_c)$. For protons remaining in the nebula, the spectrum above this energy is modified by the energy-dependent escape and becomes softer than the injection spectrum. On the contrary, the spectrum below this energy simply follows the injection spectrum. We therefore expect a softening in the proton spectrum beyond the energy $E_c=E_0(r_{\rm crab}^2/4\tau_{\rm crab}D_0)^{1/\delta}$, and subsequently a corresponding softening in the pionic gamma-ray spectrum beyond $\approx 0.1E_c$. We calculate the expected spectrum of the hadronic emission from the proton spectrum given by Eq.~\ref{eq:npspec}, following the semi-analytical method developed by \citet{Kafexhiu14}.  The attenuation to gamma-ray photons by the pair-production process on CMB is taken into account. We consider all the possible diffusion coefficients shown in Fig.~\ref{fig:diff}, and normalize the resulting 1\,PeV gamma-ray flux to $10^{-13}\, \rm erg~cm^{-2}s^{-1}$ as constrained by the LHAASO observation. As such, a soft injection proton spectrum (e.g., $\alpha>2$) is not favored as it would over-produce the measured gamma-ray spectrum below PeV.

\begin{figure}[htbp]
\includegraphics[width=1\columnwidth]{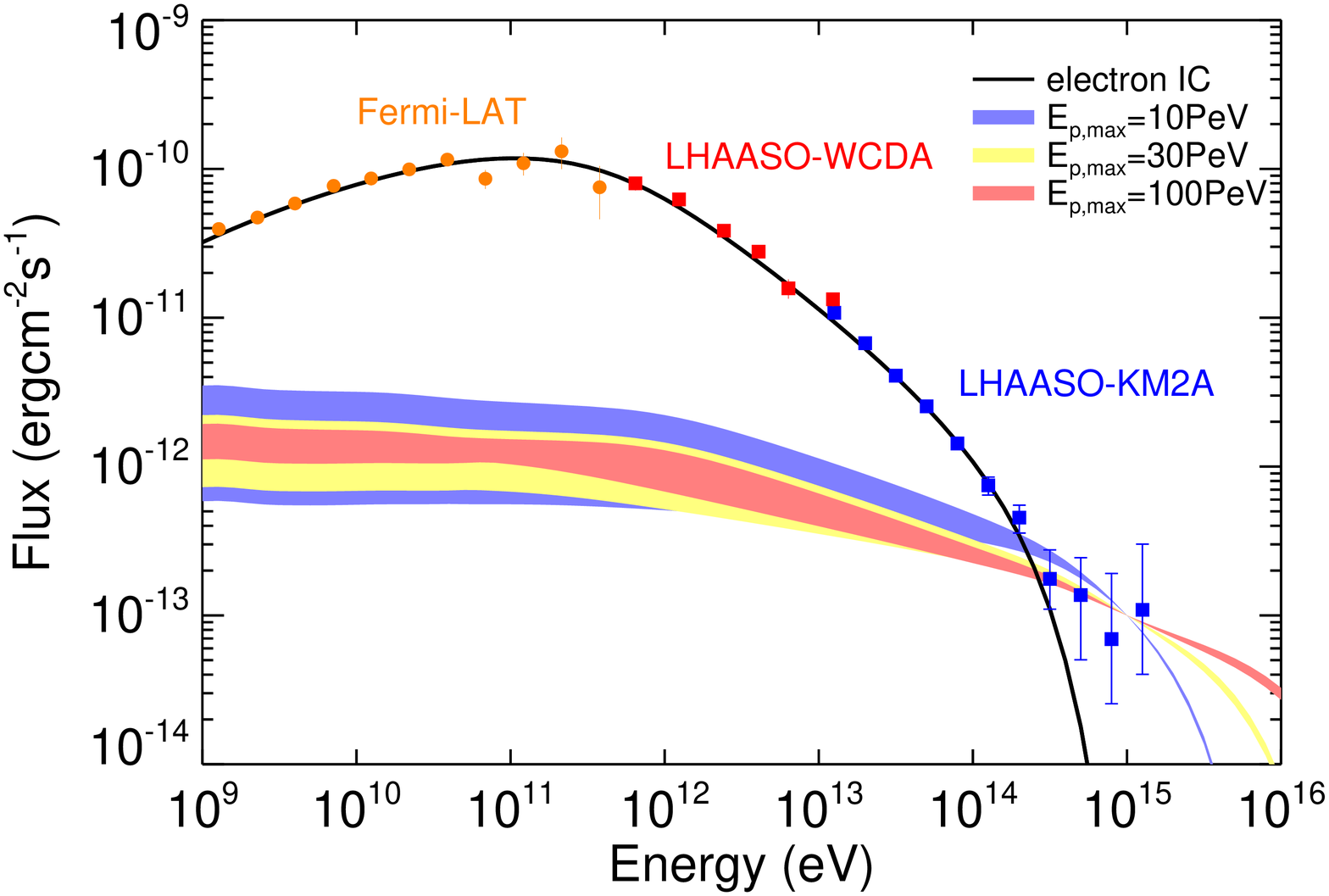}
\includegraphics[width=1\columnwidth]{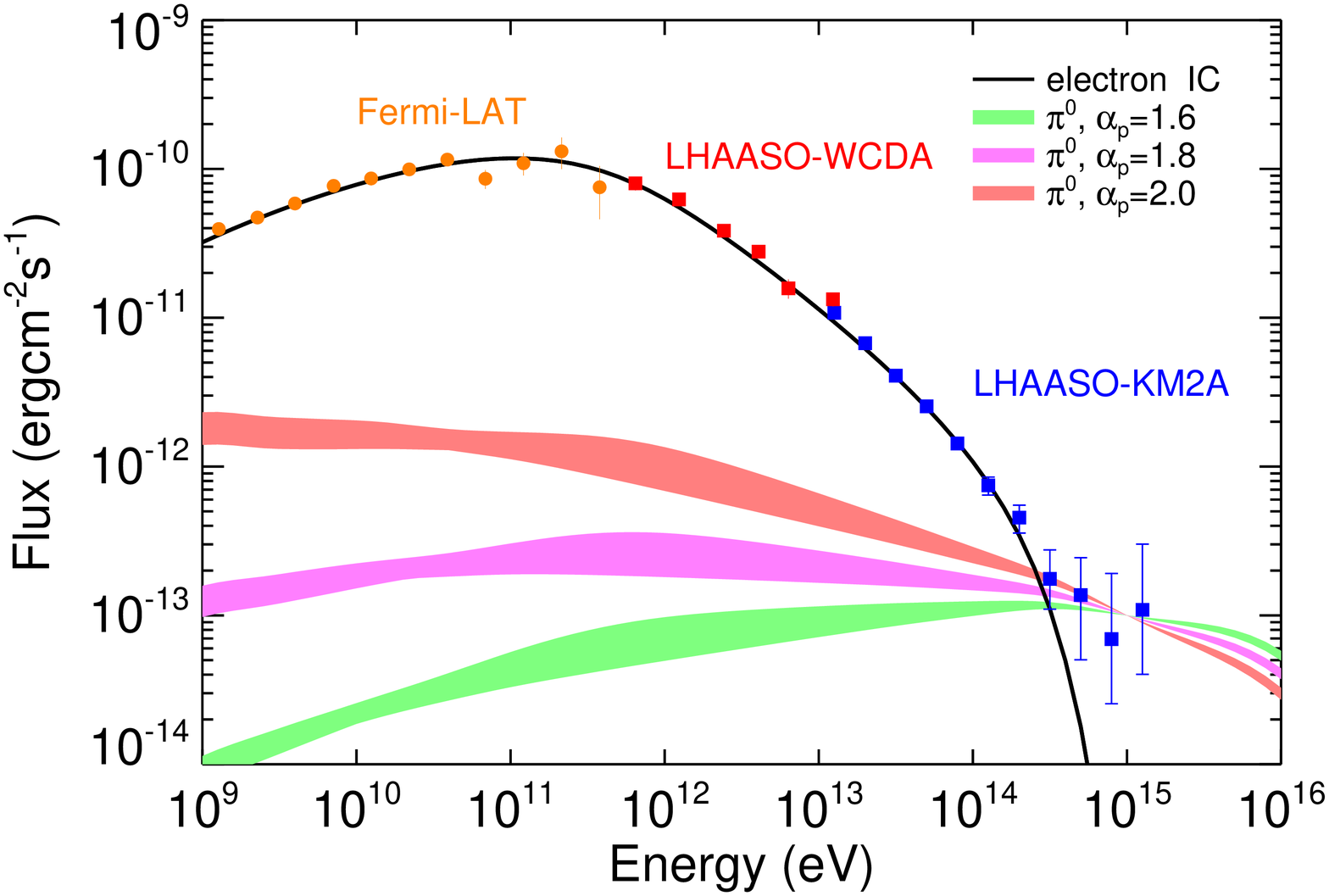}
\caption{{\bf Upper panel:} Possible hadronic gamma-ray components for a fixed proton injection spectral index of $\alpha_p=2$ for the Crab nebula. The blue, yellow, red bands represent the cases of $E_{p,\rm max}=10\,$PeV, 30\,PeV, and 100\,PeV with $\eta_p=(10-57)\%$, $(13-39)\%$, and $(21-36)\%$, respectively. The  solid black curve represents the electron  IC flux, obtained by LHAASO Collaboration (2021) for the multiwavelength emission modelling. {\bf Lower panel:} Similar to the upper panel but for a fixed maximum proton energy of $E_{p,\rm max}=100\,$PeV. The green, magenta and red bands show the cases of $\alpha_p=1.6$, 1.8 and 2.0 with $\eta_p=(8-14)\%$, $(11-18)\%$, and $(26-43)\%$, respectively. The Fermi-LAT data is taken from \citet{Arakawa20} while the LHAASO data is taken from \citet{LHAASO21_sci}. The setup of the electron component as well as the electromagnetic environment are the same with those in Figure S4 or S5 in \citet{LHAASO21_sci}, i.e.,  assuming a spectral of $\frac{dN}{dE_e}\propto E_e^{-\alpha}\left[1+(E_e/E_b)^{-\Delta \alpha}\right]^{-1}\exp\left[-(E_e/E_0)^2\right]$ where $\alpha=3.42$, $\Delta \alpha=1.76$, $E_b=0.76\,$TeV, $E_0=450$\,TeV. The target radiation field includes the synchrotron radiation in a magnetic field of $B=112\,\mu$G, a dust emission of 70\,K with an energy density of $0.5\,\rm eV~cm^{-3}$, the interstellar radiation field and the CMB.}
\label{fig:hadronic}
\end{figure}

The results are shown in Fig.~\ref{fig:hadronic} for a spectral index of $\alpha_p=2$ but with different maximum energies of accelerated protons (the upper panel) and for a fixed maximum energy of protons at 100\,PeV but with different spectral indexes $\alpha_p$ (the lower panel). A leptonic component accounting for the multiwavelength emission is also shown with employing the same parameters given by \citet{LHAASO21_sci} in their leptonic-hadronic scenario. The hadronic emission component is harder than the leptonic emission above PeV even if an extremely large $E_{e,max}$ is invoked.  Thus we can diagnose the presence of the hadronic component with the measured gamma-ray spectrum above PeV. Currently,  the large statistical uncertainties of the LHAASO measurement  at PeV  energies do not allow a strong statement about
the spectral hardening. In a more conservative way, the LHAASO data in the Figures can be regarded as the upper limit of the contribution from the possible hadronic component. This translates to an upper limit on the total energy in relativistic protons from the pulsar wind (i.e., $\eta_p$). In the case of the proton spectral slope being fixed at $\alpha_p=2.0$, we found $\eta_p=(10-57)\%$, $(13-39)\%$, and $(21-36)\%$ for $E_{p,\rm max}=10\,$PeV, 30\,PeV and 100\,PeV respectively. In the case of the maximum proton energy being fixed at $E_{p,\rm max}=100\,$PeV, we found $\eta_p=(8-14)\%$, $(11-18)\%$, and $(26-43)\%$ for $\alpha_p=1.6$, 1.8 and 2.0, respectively. The range of $\eta_p$ is due to the uncertainty of the diffusion coefficient considered in the calculation. In general, we see that the current LHAASO data allows a fraction of $10\%-50\%$ spindown energy to be converted to the relativistic proton energy in our model.

The above discussions have assumed that the hadronic component in the pulsar wind is pure protons. However, the composition is not well-known and heavier nuclei could show up in the nebula if the photo-disintegration effect is not important. For nuclei with a mass number $A$, the cross section of the hadronuclear interaction becomes $A$ times larger but the energy fraction lost into created pions in one collision is also reduced by $1/A$. As a result, even if heavier nuclei are presented in the nebula, it would not change the above result significantly.

\section{Discussions}
\subsection{Possible Contribution to Cosmic Rays above the Knee}
In our scenario, a considerable fraction of Crab pulsar's spindown energy could be converted into relativistic protons, but most of them have already escape the Crab Nebula, becoming cosmic rays. Considering that the age of the pulsar is about 1000 years and it is located at 6500 light years ($\simeq 2\,$kpc) away from Earth, protons that escape at the earliest epoch have travelled a distance of $2\sqrt{D_{\rm ISM}\times 7500\rm yr}$. Assuming the cosmic-ray diffusion coefficient in the ISM $D_{\rm ISM}\simeq 4\times10^{28}(E/1\,{\rm GeV})^{1/3}\rm cm^2s^{-1}$ \citep[e.g.][]{Trotta11}, those protons can travel to a distance of $l=0.9(E_p/10{\rm PeV})^{1/6}\,$kpc, and have not arrived at Earth. On the other hand, if all the pulsars are cosmic-ray proton accelerators, they might contribute to the measured cosmic ray flux. According to the ATNF Pulsar Catalogue \citep{Manchester05}, the sum of spindown luminosities of Galactic pulsars is about $L_{\rm s, tot}\approx 10^{39}\rm erg~s^{-1}$. Assuming that such a luminosity is stable over million years and {that $\eta_p$ is the same for all pulsars}\footnote{Using the same $\eta_p$ for all pulsars is a naive assumption.   The nuclei extracted from the surface of a neutron star are accelerated, at first when passing the outer gap of the inner pulsar magnetosphere and later in the pulsar wind. Due to the lack of knowledge about the details of the above processes, we follow previous assumption that a same fraction of  the pulsar rotational energy is lost on acceleration of iron nuclei (and then to relativistic protons due to photo-disintegration) for all galactic pulsars \citep{Bednarek97,Bednarek04}.}, we may roughly estimate the flux of cosmic rays produced by pulsars, in the framework of the leaky box model, by
\begin{equation}
\begin{split}
 F(E_p)&=\frac{c}{4\pi}\frac{\eta_p L_{\rm s,tot}t_{\rm esc}}{2\pi R_{\rm Gal}^2H_{\rm CR}\ln (E_{p,\rm max}/E_{p,\rm min})}\\
 &\approx 2\times 10^3 \left(\frac{f_{\rm pul}\eta_pL_{\rm s, tot}}{10^{39}\,\rm erg~s^{-1}}\right)\left(\frac{E_p}{10\,{\rm PeV}}\right)^{-1/3}\\
 &\times \left(\frac{H_{\rm CR}}{4\,\rm kpc}\right)\left(\frac{R_{\rm Gal}}{15\,\rm kpc}\right)^{-2}\, \rm eV~cm^{-2}s^{-1}sr^{-1}
\end{split}
\end{equation}
where $t_{\rm esc}=H_{\rm CR}^2/4D_{\rm ISM}$ is the cosmic-ray residence time in the Galaxy with $H_{\rm CR}$ being the scale height of the cosmic-ray halo. $R_{\rm Gal}$ is the radius of the Galaxy. $f_{\rm pul}(\geq 1)$ accounts for the contribution of off-beamed pulsars that we cannot observe, and its value could be a factor of a few according to the study by \citet{Tauris98}. Here we have assumed a flat proton spectrum with $\ln (E_{p,\rm max}/E_{p,\rm min})\simeq 20$. The KASCADE-Grande experiment has revealed a cosmic-ray proton flux of $\sim 3000 \,\rm eV~cm^{-2}s^{-1}sr^{-1}$ at 10\,PeV and $\sim 100 \,\rm eV~cm^{-2}s^{-1}sr^{-1}$ at 100\,PeV \citep{KASCADE13b}. It implies that Galactic pulsars may have a significant contribution to the cosmic-ray proton flux between $10-100\,$PeV provided an approximate choice of the value of $\eta_p$, which is consistent with the previous suggestion \citep{Bednarek04}.

\subsection{The Second Leptonic Component}
The spectral hardening of the Crab Nebula at PeV energies could also be due to a second electron component. \citet{LHAASO21_sci} studied such a scenario with the assumption of a Maxwellian-type spectrum \citep[see also][]{Aharonian98} and an $E^{-1.5}$ spectrum for the second electron component. We here explore more general cases with the hard electron spectrum, considering the maximum energy $E_{e,\rm max}$ and the slope of the electron spectrum $\alpha$ as free parameters to study their influences, where the electron spectrum is defined as  $dN_e/dE_e\propto E_e^{-\alpha}{\rm exp}(-E_e/E_{e,\rm max})$ (for details in Appendix.~\ref{sec:app}).

The results of the spectra are shown in Fig.~\ref{fig:KN}.  We find that, although the electron IC radiation suffers the KN effect at such high energies, it could still account for a spectral hardening up to several PeV (or even higher) energy provided that the electron spectrum is sufficiently hard and the maximum energy is sufficiently high. The critical point is that whether the required maximum electron energy and the hard spectrum can be generated. The former one can be determined by the balance between the acceleration and the synchrotron energy lose rates, which reads \citep{LHAASO21_sci}
\begin{equation}\label{eq:emax}
E_{e,\rm max}=6 \left(\frac{B}{100\mu G}\right)^{-1/2}{\rm PeV},
\end{equation}
in the most efficient acceleration case. Although $B\approx 100\mu$G is obtained via the SED fitting in the one-zone model, the magnetic field strength inside the nebula could be weaker at the termination shock  (i.e., the acceleration site) than at larger distance \citep{KC84a, Atoyan96}. Thus, acceleration of $\sim 10\,$PeV electrons in the nebula may still be possible. On the other hand, the hard spectrum may be achieved in the magnetic reconnection process in optimistic conditions \citep{Sironi14, Guo14, Werner16}. Therefore, it may not be easy to distinguish between the hadronic and the leptonic origin for the spectral hardening (if confirmed) solely from the UHE spectrum measurement. Future sensitive neutrino telescopes (e.g. IceCube Gen-2) might be able to distinguish between the leptonic and hadronic scenarios for the spectral hardening \citep{Aartsen2021}, since the hadronic scenario predict a neutrinos flux comparable to the PeV gamma-ray flux.

\section{Summary}
To summarize, based on the latest measurement of PeV emission of the Crab Nebula by LHAASO,  we put a constraint on the proton content accelerated in pulsar wind by taking into account the particle escape effect. We found that the current LHAASO data allows up to $(10-50)\%$ of the spindown energy being converted to relativistic protons. This is much larger that previous limits obtained without considering the diffusive escape effect \citep{ZhangX20}. Future observations with the full LHAASO will be able to determine reliably whether there is a hard component in the spectrum of the Crab Nebula. If this hadronic component is confirmed in future,  our results may imply that  a significant fraction of spindown power is converted into relativistic ions in the pulsar wind. In addition, Galactic pulsars  may make a considerable contribution to the cosmic-ray proton flux at $10-100$\,PeV provided the proton spectrum is not too soft. A second electron component might also explain the spectral hardening beyond PeV energy provided a sufficiently high maximum electron energy and a sufficiently hard spectrum. It'd therefore be difficult to distinguish the hadronic and leptonic origins for the spectral hardening (if confirmed in the future) solely from the spectrum measurement, but the future neutrino observation could provide an important test.

\section*{Acknowledgments}
We thank Felix Aharonian for invaluable discussions. This work is supported by the National Key R \& D program of China under the grant 2018YFA0404203 and the NSFC  grants
11625312, 11851304 and U2031105.

\newpage
\appendix
\section{Electron Inverse Compton radiation at Ultrahigh Energy}\label{sec:app}

We consider the secondary electron spectrum in the PWN to be a power-law function with an exponential cutoff, $dN_e/dE_e\propto E_e^{-\alpha}{\rm exp}(-E_e/E_{e,\rm max})$. Since we are interested in the spectrum of gamma-rays above 100\,TeV, we only consider the CMB as the target photons, neglecting the contribution by the dust emission and the synchrotron radiation of electrons due to the severe KN suppression effect.
The maximum energy $E_{e,\rm max}$ may depend on the strength of the magnetic field in the particle acceleration zone \citep{LHAASO21_sci}, but here we consider it as free parameter to study its influence.

\begin{figure*}[htbp]
\centering
\includegraphics[width=0.45\textwidth]{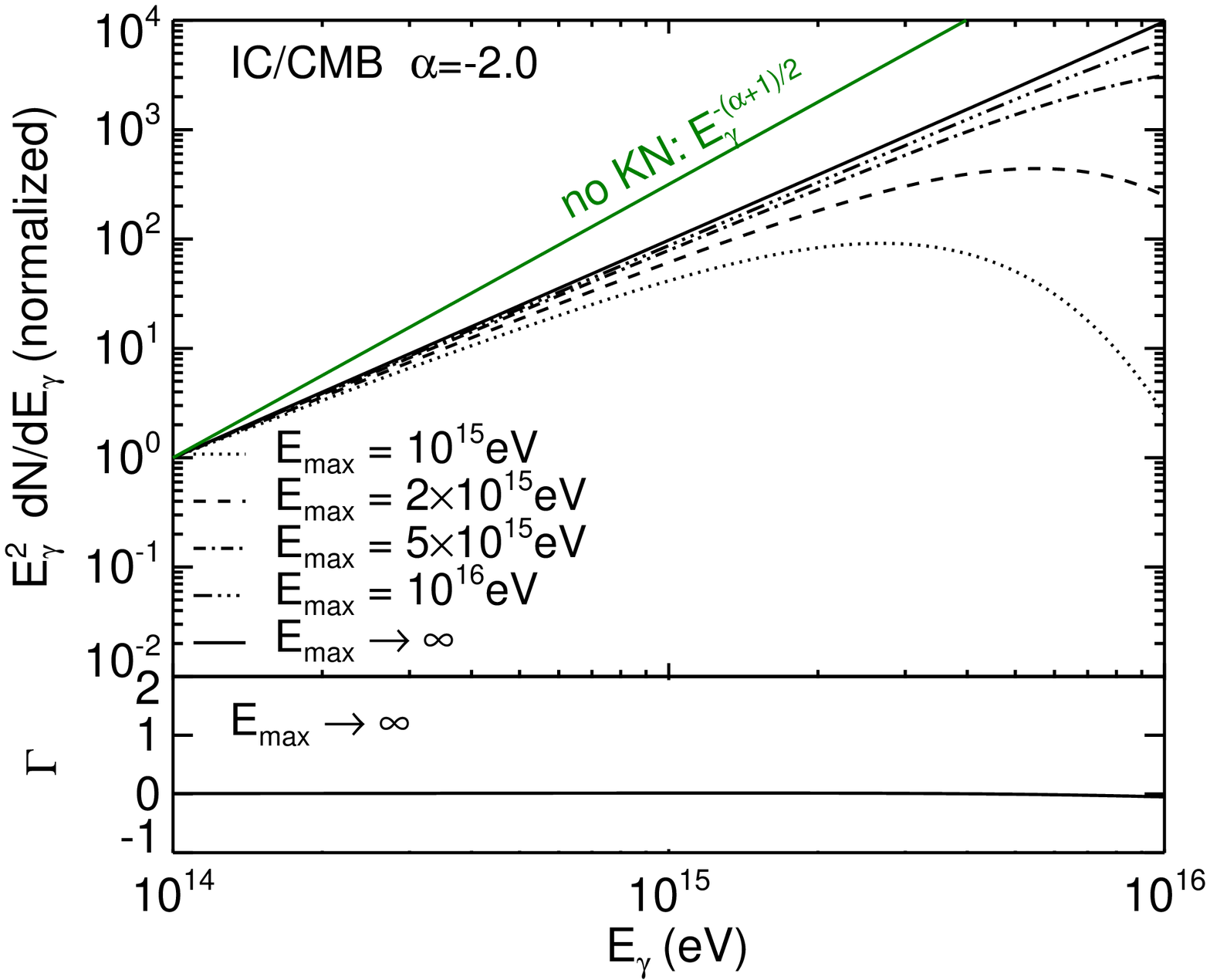}
\includegraphics[width=0.45\textwidth]{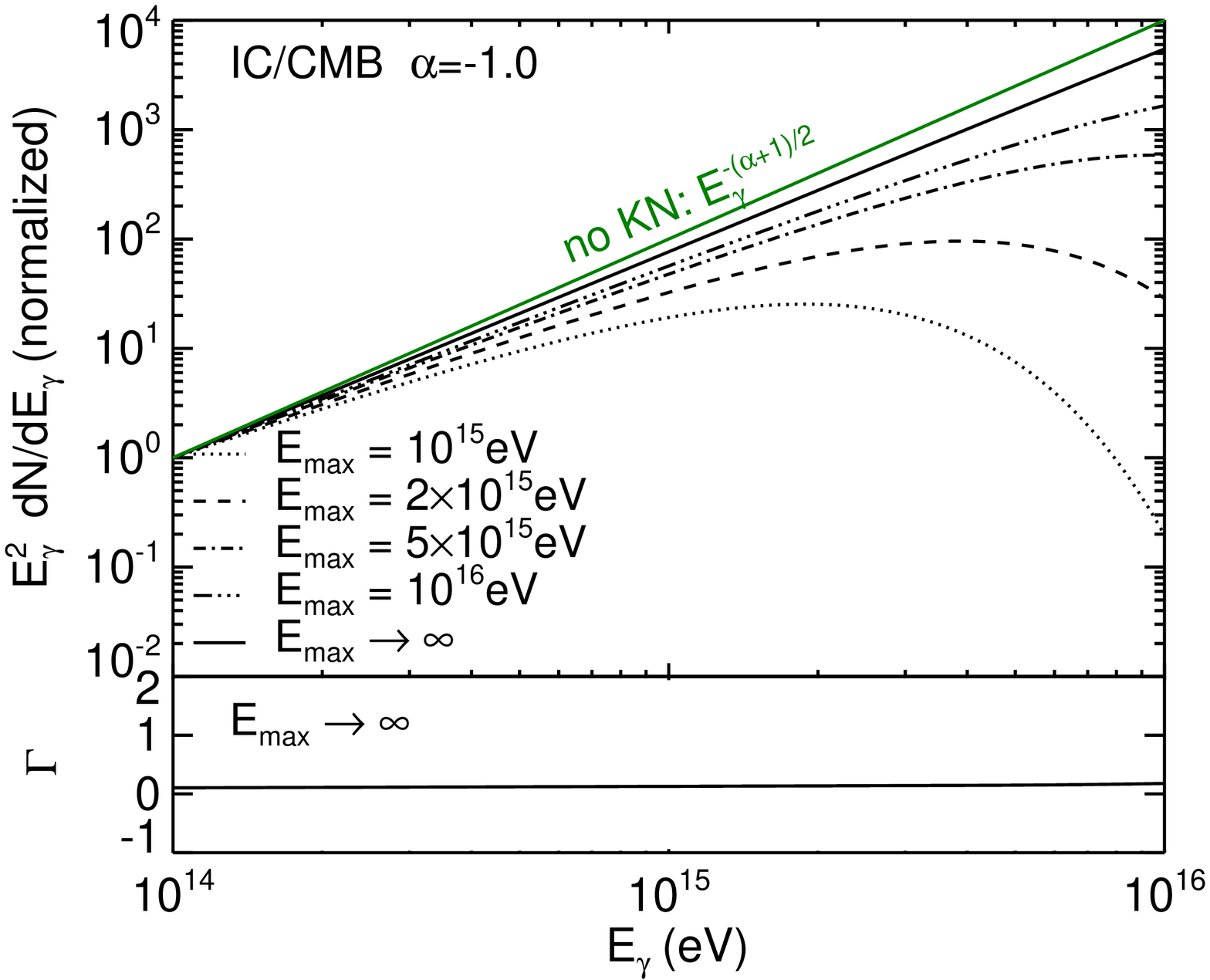}
\includegraphics[width=0.45\textwidth]{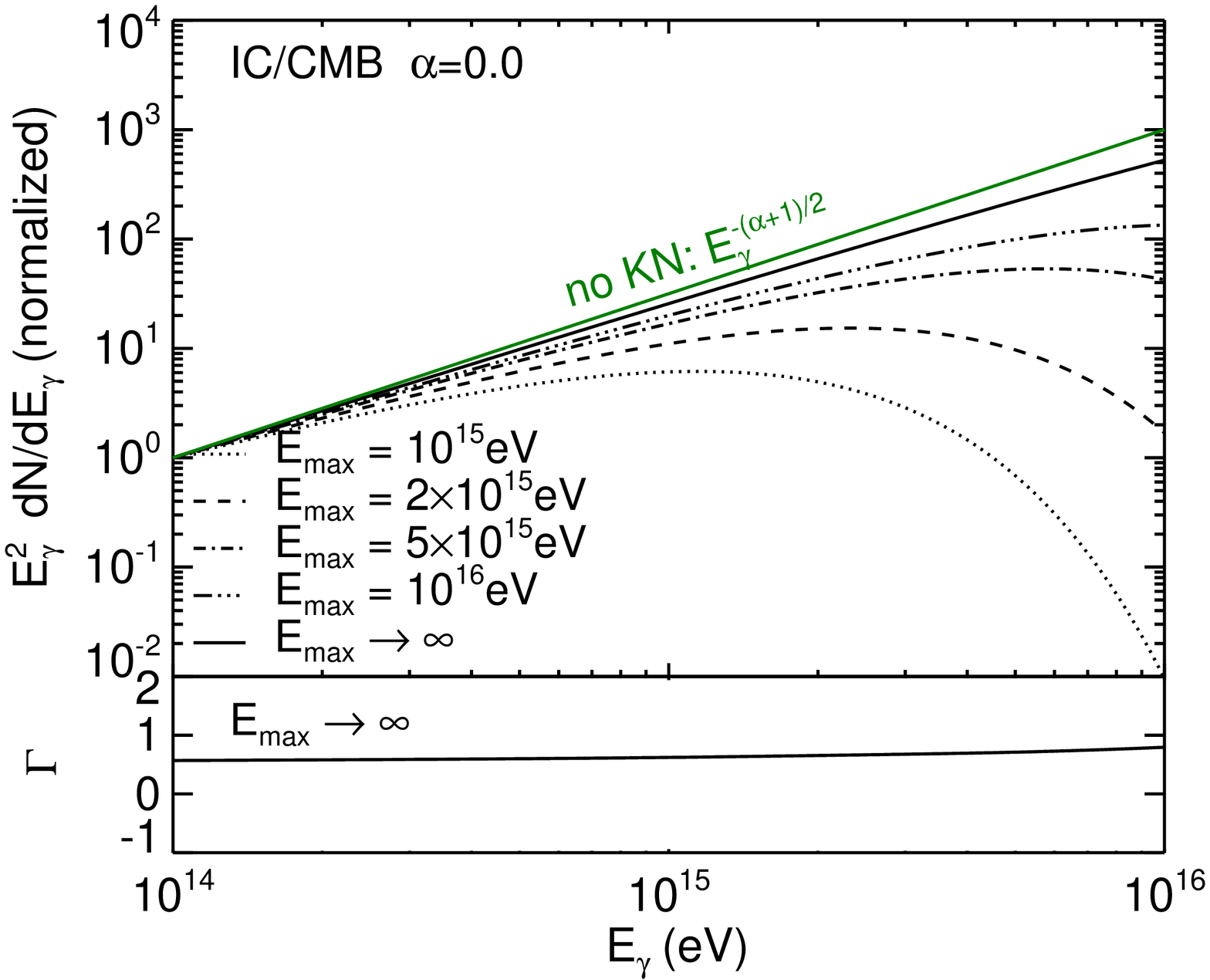}
\includegraphics[width=0.45\textwidth]{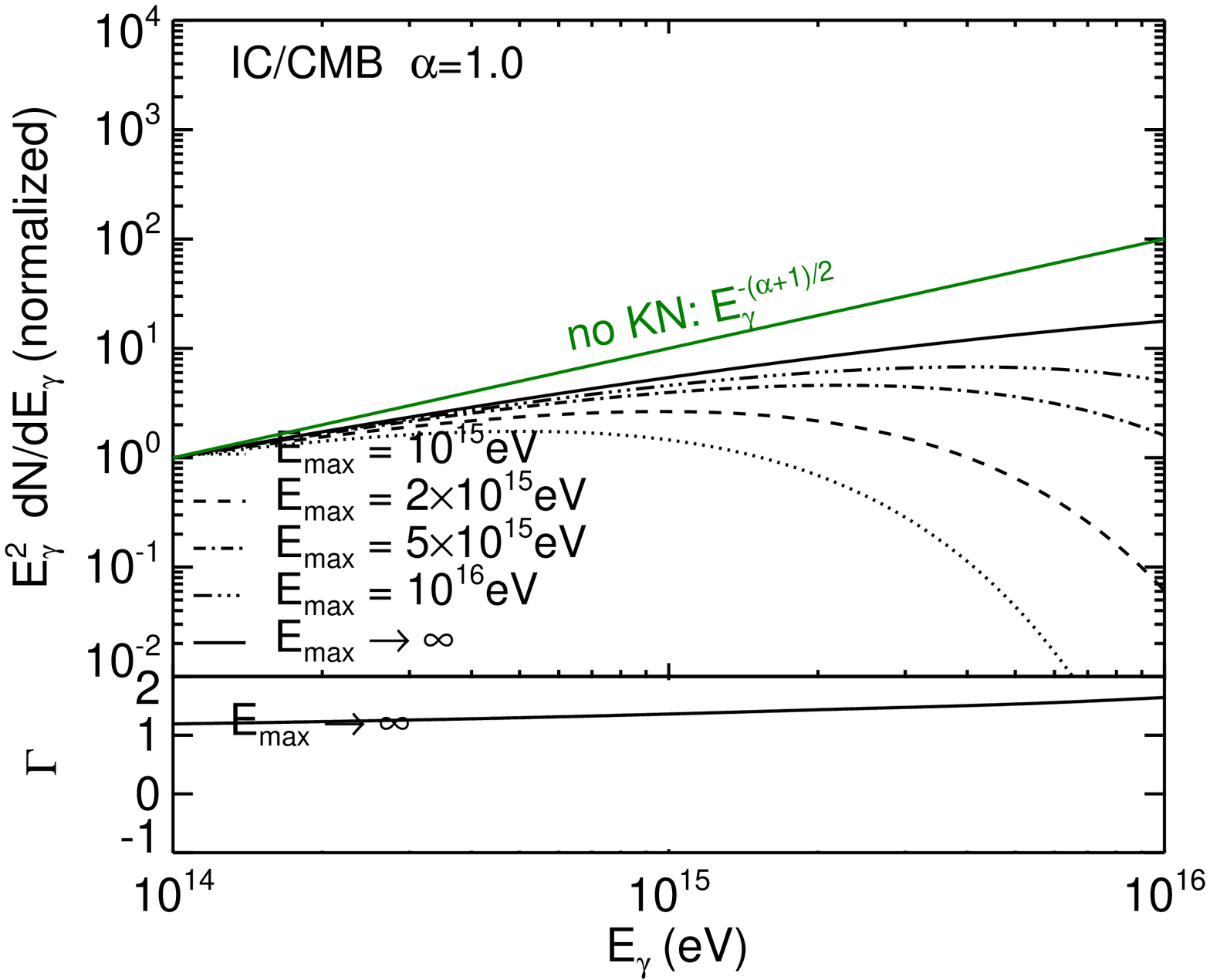}
\caption{Gamma-ray spectra via IC scattering off CMB and spectral indexes for electrons of power-law distribution with a super-exponential cutoff, with the power-law index being $\alpha=-2$ or Maxwellian distribution ({\bf top-left panel}), $\alpha=-1$ ({\bf top-right panel}), $\alpha=0$ ({\bf bottom-left panel}) and $\alpha=1$ ({\bf bottom-right panel}). The dotted, dashed, dot-dashed, dot-dot-dot-dashed black curves represent the cases with maximum electron energy of $E_{e,\rm max}=10^{15}\,$eV, $2\times 10^{15}\,$eV, $5\times 10^{15}\,$eV and $10^{16}\,$eV respectively, while the solid black curves present the limiting case with $E_{e,\rm max}\to \infty$. The solid green curves show the expected IC spectra without the KN effect for reference.}
\label{fig:KN}
\end{figure*}

Since the information of the target photons (i.e., the CMB) for the IC emission is known, the IC spectrum is only dependent of the electron spectral slope $\alpha$ and the maximum energy. For a hard electron spectrum, we here consider $\alpha=-2, -1, 0, 1$ and show the scaled IC spectrum in Fig.~\ref{fig:KN} with considering different maximum energy $E_{e,\rm max}$. Note that the spectrum with $\alpha=-2$ is the Maxwellian-type distribution. The local photon indexes $\Gamma$ (defined as $dN_\gamma/dE_\gamma \propto E_\gamma^{-\Gamma(E_\gamma)}$) with $E_{e,\rm max}\to \infty$ are shown in the corresponding lower panels.
As can be seen from Fig.~\ref{fig:KN}, the IC spectrum is softer than the case without considering the KN effect. However, with a hard electron spectrum, the resulting gamma-ray spectrum can be quite flat or even harder. This could be also understood analytically. Bearing in mind that $E_\gamma\sim E_e$ in the deep KN regime, the IC energy spectrum can be approximated by
\begin{equation}
L_{\gamma, \rm IC}(E_\gamma)\approx \frac{E_e^2dN_e/dE_e}{t_{\rm IC}(E_e)} \propto \frac{E_e^{2-\alpha}}{E_e^{0.7}}\propto E_\gamma^{1.3-\alpha}.
\end{equation}
Here $t_{\rm IC}(E_e)\propto E_e^{0.7}$ is the cooling timescale of electrons via the IC process in the deep KN regime \citep{Khangulyan14}. We then can find the photon index $\Gamma=\alpha+0.7$. On the other hand, in the deep KN regime, the IC spectrum at the low-energy limit (i.e, $E_\gamma \ll E_e$) is $dN_\gamma/dE_\gamma \propto E_\gamma^0$ for a single IC scattering event\footnote{According to \citet{Blumenthal70}, the distribution of IC scattered photon spectrum can be given by $F(E_\gamma)=2q\ln q+(1+2q)(1-q)+(1-q)(\Gamma_\epsilon q)^2/2(1+\Gamma_\epsilon q)$, where $\Gamma_\epsilon=4\epsilon\gamma/m_ec^2$, $q=E_1/\Gamma_\epsilon(1-E_1)$, and $E_1=E_\gamma/E_e$. In the deep KN regime ($\Gamma_\epsilon\gg 1$) and the low-energy limit ($E_1 \ll 1$), we have $q\ll 1$ and $\Gamma_\epsilon q \ll 1$, leading to $F(E_\gamma)\to 1$.} . As a result, the low-energy IC spectrum would be dominated by the low-energy tail emitted by electrons around the cutoff energy $E_{e,\rm max}$ if $\alpha < -0.7$, leading to $\Gamma \approx 0$, which is consistent with the results shown in the panels for $\alpha=-2$ and $\alpha=-1$.

%
%

\bibliographystyle{apj}
\bibliography{ms}

\begin{thebibliography}{44}
\expandafter\ifx\csname natexlab\endcsname\relax\def\natexlab#1{#1}\fi

\bibitem[{{Aharonian} \& {Atoyan}(1998)}]{Aharonian98}
{Aharonian}, F.~A. \& {Atoyan}, A.~M. 1998, NewAR, 42, 579

\bibitem[{{Aloisio} {et~al.}(2009){Aloisio}, {Berezinsky}, \&
  {Gazizov}}]{Aloisio09}
{Aloisio}, R., {Berezinsky}, V., \& {Gazizov}, A. 2009, \apj, 693, 1275

\bibitem[{{Amenomori} {et~al.}(2019){Amenomori}, {Bao}, {Bi}, {Chen}, {Chen},
  {Chen}, {Chen}, {Chen}, {Cirennima}, {Cui}, {Danzengluobu}, {Ding}, {Fang},
  {Fang}, {Feng}, {Feng}, {Feng}, {Gao}, {Gou}, {Guo}, {He}, {He}, {Hibino},
  {Hotta}, {Hu}, {Hu}, {Huang}, {Jia}, {Jiang}, {Jin}, {Kajino}, {Kasahara},
  {Katayose}, {Kato}, {Kato}, {Kawata}, {Kozai}, {Labaciren}, {Le}, {Li}, {Li},
  {Li}, {Lin}, {Liu}, {Liu}, {Liu}, {Liu}, {Lou}, {Lu}, {Meng}, {Mitsui},
  {Munakata}, {Nakamura}, {Nanjo}, {Nishizawa}, {Ohnishi}, {Ohta}, {Ozawa},
  {Qian}, {Qu}, {Saito}, {Sakata}, {Sako}, {Sengoku}, {Shao}, {Shibata},
  {Shiomi}, {Sugimoto}, {Takita}, {Tan}, {Tateyama}, {Torii}, {Tsuchiya},
  {Udo}, {Wang}, {Wu}, {Xue}, {Yagisawa}, {Yamamoto}, {Yang}, {Yuan}, {Zhai},
  {Zhang}, {Zhang}, {Zhang}, {Zhang}, {Zhang}, {Zhang}, {Zhang},
  {Zhaxisangzhu}, {Zhou}, \& {Tibet AS {\ensuremath{\gamma}}
  Collaboration}}]{ASgamma19}
{Amenomori}, M. {et al.}\  2019, \prl, 123, 051101

\bibitem[{{Apel} {et~al.}(2013){Apel}, {Arteaga-Vel{\'a}zquez}, {Bekk},
  {Bertaina}, {Bl{\"u}mer}, {Bozdog}, {Brancus}, {Cantoni}, {Chiavassa},
  {Cossavella}, {Daumiller}, {de Souza}, {Di Pierro}, {Doll}, {Engel},
  {Engler}, {Finger}, {Fuchs}, {Fuhrmann}, {Gils}, {Glasstetter}, {Grupen},
  {Haungs}, {Heck}, {H{\"o}randel}, {Huber}, {Huege}, {Kampert}, {Kang},
  {Klages}, {Link}, {{\L}uczak}, {Ludwig}, {Mathes}, {Mayer}, {Melissas},
  {Milke}, {Mitrica}, {Morello}, {Oehlschl{\"a}ger}, {Ostapchenko}, {Palmieri},
  {Petcu}, {Pierog}, {Rebel}, {Roth}, {Schieler}, {Schoo}, {Schr{\"o}der},
  {Sima}, {Toma}, {Trinchero}, {Ulrich}, {Weindl}, {Wochele}, {Wommer}, \&
  {Zabierowski}}]{KASCADE13b}
{Apel}, W.~D. {et al.}\  2013, Astroparticle Physics, 47, 54

\bibitem[{{Arakawa} {et~al.}(2020){Arakawa}, {Hayashida}, {Khangulyan}, \&
  {Uchiyama}}]{Arakawa20}
{Arakawa}, M., {Hayashida}, M., {Khangulyan}, D., \& {Uchiyama}, Y. 2020, \apj,
  897, 33

\bibitem[Aartsen et al. (2021)]{Aartsen2021} {Aartsen}, M.~G., {Abbasi}, R. , {Ackermann}, M, et al.
2021, Journal of Physics G Nuclear Physics, 48, 060501

\bibitem[{{Arons} \& {Tavani}(1994)}]{Arons94}
{Arons}, J. \& {Tavani}, M. 1994, \apjs, 90, 797

\bibitem[{{Atoyan} \& {Aharonian}(1996)}]{Atoyan96}
{Atoyan}, A.~M. \& {Aharonian}, F.~A. 1996, \mnras, 278, 525

\bibitem[{{Atoyan} {et~al.}(1995){Atoyan}, {Aharonian}, \&
  {V{\"o}lk}}]{Atoyan95}
{Atoyan}, A.~M., {Aharonian}, F.~A., \& {V{\"o}lk}, H.~J. 1995, \prd, 52, 3265

\bibitem[{{Bednarek} \& {Bartosik}(2004)}]{Bednarek04}
{Bednarek}, W. \& {Bartosik}, M. 2004, \aap, 423, 405

\bibitem[{{Bednarek} \& {Protheroe}(1997)}]{Bednarek97}
{Bednarek}, W. \& {Protheroe}, R.~J. 1997, \prl, 79, 2616

\bibitem[{{Blumenthal} \& {Gould}(1970)}]{Blumenthal70}
{Blumenthal}, G.~R. \& {Gould}, R.~J. 1970, Reviews of Modern Physics, 42, 237

\bibitem[{{Cheng} {et~al.}(1990){Cheng}, {Cheung}, {Lau}, {Yu}, \&
  {Kwok}}]{ChengKS90}
{Cheng}, K.~S., {Cheung}, T., {Lau}, M.~M., {Yu}, K.~N., \& {Kwok}, P.~W. 1990,
  Journal of Physics G Nuclear Physics, 16, 1115

\bibitem[{{Cheng} {et~al.}(1986){Cheng}, {Ho}, \& {Ruderman}}]{ChengKS86}
{Cheng}, K.~S., {Ho}, C., \& {Ruderman}, M. 1986, \apj, 300, 500

\bibitem[{{de Jager} \& {Djannati-Ata{\"\i}}(2009)}]{deJager09}
{de Jager}, O.~C. \& {Djannati-Ata{\"\i}}, A. 2009, in Astrophysics and Space
  Science Library, Vol. 357, Astrophysics and Space Science Library, ed.
  W.~{Becker}, 451

\bibitem[{{de Jager} \& {Harding}(1992)}]{deJager92}
{de Jager}, O.~C. \& {Harding}, A.~K. 1992, \apj, 396, 161

\bibitem[{{Gallant} \& {Arons}(1994)}]{Gallant94}
{Gallant}, Y.~A. \& {Arons}, J. 1994, \apj, 435, 230

\bibitem[{{Goldreich} \& {Julian}(1969)}]{GJ69}
{Goldreich}, P. \& {Julian}, W.~H. 1969, \apj, 157, 869

\bibitem[{{Guo} {et~al.}(2014){Guo}, {Li}, {Daughton}, \& {Liu}}]{Guo14}
{Guo}, F., {Li}, H., {Daughton}, W., \& {Liu}, Y.-H. 2014, \prl, 113, 155005

\bibitem[{{Horns} {et~al.}(2006){Horns}, {Aharonian}, {Santangelo}, {Hoffmann},
  \& {Masterson}}]{Horns06}
{Horns}, D., {Aharonian}, F., {Santangelo}, A., {Hoffmann}, A.~I.~D., \&
  {Masterson}, C. 2006, \aap, 451, L51

\bibitem[{{Hoshino} {et~al.}(1992){Hoshino}, {Arons}, {Gallant}, \&
  {Langdon}}]{Hoshino92}
{Hoshino}, M., {Arons}, J., {Gallant}, Y.~A., \& {Langdon}, A.~B. 1992, \apj,
  390, 454

\bibitem[{{Kafexhiu} {et~al.}(2014){Kafexhiu}, {Aharonian}, {Taylor}, \&
  {Vila}}]{Kafexhiu14}
{Kafexhiu}, E., {Aharonian}, F., {Taylor}, A.~M., \& {Vila}, G.~S. 2014, \prd,
  90, 123014

\bibitem[{{Kennel} \& {Coroniti}(1984{\natexlab{a}})}]{KC84b}
{Kennel}, C.~F. \& {Coroniti}, F.~V. 1984{\natexlab{a}}, \apj, 283, 694

\bibitem[{{Kennel} \& {Coroniti}(1984{\natexlab{b}})}]{KC84a}
--- 1984{\natexlab{b}}, \apj, 283, 710

\bibitem[{{Khangulyan} {et~al.}(2014){Khangulyan}, {Aharonian}, \&
  {Kelner}}]{Khangulyan14}
{Khangulyan}, D., {Aharonian}, F.~A., \& {Kelner}, S.~R. 2014, \apj, 783, 100

\bibitem[{{LHAASO Collaboration}(2021{\natexlab{a}})}]{LHAASO21_sci}
{LHAASO Collaboration} 2021{\natexlab{a}}, \sci, 373, 425

\bibitem[{{LHAASO Collaboration}(2021{\natexlab{b}})}]{LHAASO21_nat}
--- 2021{\natexlab{b}}, \nat, 594, 33

\bibitem[{{Li} {et~al.}(2010){Li}, {Chen}, \& {Zhang}}]{LiH10}
{Li}, H., {Chen}, Y., \& {Zhang}, L. 2010, \mnras, 408, L80

\bibitem[{{Lundmark}(1921)}]{Lundmark1921}
{Lundmark}, K. 1921, \pasp, 33, 225

\bibitem[{{Lyne} {et~al.}(1988){Lyne}, {Pritchard}, \& {Smith}}]{Lyne88}
{Lyne}, A.~G., {Pritchard}, R.~S., \& {Smith}, F.~G. 1988, \mnras, 233, 667

\bibitem[{{Manchester} {et~al.}(2005){Manchester}, {Hobbs}, {Teoh}, \&
  {Hobbs}}]{Manchester05}
{Manchester}, R.~N., {Hobbs}, G.~B., {Teoh}, A., \& {Hobbs}, M. 2005, \aj, 129,
  1993

\bibitem[{{Owen} \& {Barlow}(2015)}]{Owen15}
{Owen}, P.~J. \& {Barlow}, M.~J. 2015, \apj, 801, 141

\bibitem[{{Pacini} \& {Salvati}(1973)}]{Pacini73}
{Pacini}, F. \& {Salvati}, M. 1973, \apj, 186, 249

\bibitem[{{Prosekin} {et~al.}(2015){Prosekin}, {Kelner}, \&
  {Aharonian}}]{Prosekin15}
{Prosekin}, A.~Y., {Kelner}, S.~R., \& {Aharonian}, F.~A. 2015, \prd, 92,
  083003

\bibitem[{{Rees} \& {Gunn}(1974)}]{Rees74}
{Rees}, M.~J. \& {Gunn}, J.~E. 1974, \mnras, 167, 1

\bibitem[{{Reynolds} \& {Chevalier}(1984)}]{Reynolds84}
{Reynolds}, S.~P. \& {Chevalier}, R.~A. 1984, \apj, 278, 630

\bibitem[{{Sironi} \& {Spitkovsky}(2014)}]{Sironi14}
{Sironi}, L. \& {Spitkovsky}, A. 2014, \apjl, 783, L21

\bibitem[{{Tang} \& {Chevalier}(2012)}]{Tang12}
{Tang}, X. \& {Chevalier}, R.~A. 2012, \apj, 752, 83

\bibitem[{{Tauris} \& {Manchester}(1998)}]{Tauris98}
{Tauris}, T.~M. \& {Manchester}, R.~N. 1998, \mnras, 298, 625

\bibitem[{{Trotta} {et~al.}(2011){Trotta}, {J{\'o}hannesson}, {Moskalenko},
  {Porter}, {Ruiz de Austri}, \& {Strong}}]{Trotta11}
{Trotta}, R., {J{\'o}hannesson}, G., {Moskalenko}, I.~V., {Porter}, T.~A.,
  {Ruiz de Austri}, R., \& {Strong}, A.~W. 2011, \apj, 729, 106

\bibitem[{{Weisskopf} {et~al.}(2000){Weisskopf}, {Hester}, {Tennant}, {Elsner},
  {Schulz}, {Marshall}, {Karovska}, {Nichols}, {Swartz}, {Kolodziejczak}, \&
  {O'Dell}}]{Weisskopf00}
{Weisskopf}, M.~C. {et al.}\  2000, \apjl, 536, L81

\bibitem[{{Werner} {et~al.}(2016){Werner}, {Uzdensky}, {Cerutti}, {Nalewajko},
  \& {Begelman}}]{Werner16}
{Werner}, G.~R., {Uzdensky}, D.~A., {Cerutti}, B., {Nalewajko}, K., \&
  {Begelman}, M.~C. 2016, \apjl, 816, L8

\bibitem[{{Yang} \& {Zhang}(2009)}]{YangX09}
{Yang}, X.~C. \& {Zhang}, L. 2009, \aap, 496, 751

\bibitem[{{Zhang} \& {Yang}(2009)}]{ZhangL09}
{Zhang}, L. \& {Yang}, X.~C. 2009, \apjl, 699, L153

\bibitem[{{Zhang} {et~al.}(2020){Zhang}, {Chen}, {Huang}, \& {Chen}}]{ZhangX20}
{Zhang}, X., {Chen}, Y., {Huang}, J., \& {Chen}, D. 2020, \mnras, 497, 3477

\end{thebibliography}

\end{document}